\documentclass[twocolumn,preprintnumbers,superscriptaddress,endnote,nofootinbib,aps,prd,floatfix]{revtex4}

\usepackage{footmisc,multirow}
\usepackage{subfigure}
\usepackage{amsmath,slashed}

\usepackage{graphicx}

\newcommand{\be}{\begin{eqnarray*}}
\newcommand{\ee}{\end{eqnarray*}}
\newcommand{\gl}[1]{(\ref{#1})}

\newcommand{\bee}{\begin{eqnarray}}
\newcommand{\eee}{\end{eqnarray}}
\newcommand{\beeq}{\begin{equation}}
\newcommand{\eeeq}{\end{equation}}
\newcommand{\gev}{{\rm{GeV}}}
\newcommand{\tev}{{\rm{TeV}}}
\newcommand{\fb}{{\rm{fb}}}

\newcommand{\pb}{{\rm{pb}}}

\begin{document}

\title{Higgs self-coupling measurements at the LHC}

\begin{abstract}
  Both the ATLAS and CMS collaborations have reported a Standard Model
  Higgs-like excess at around $m_h = 125$~GeV. If an SM-like Higgs
  particle is discovered in this particular mass range, an important
  additional test of the SM electroweak symmetry breaking sector is
  the measurement of the Higgs self-interactions. We investigate the
  prospects of measuring the Higgs self-coupling for $m_h=125~\gev$ in
  the dominant SM decay channels in boosted and unboosted kinematical
  regimes.  We further enhance sensitivity by considering dihiggs
  systems recoiling against a hard jet. This configuration exhibits a
  large sensitivity to the Higgs self-coupling which can be accessed
  in subjet-based analyses. Combining our analyses allows constraints
  to be set on the Higgs self-coupling at the LHC.
\end{abstract}

\author{Matthew J. Dolan} 
\affiliation{Institute for Particle Physics Phenomenology, Department
  of Physics,\\Durham University, DH1 3LE, United Kingdom}
\author{Christoph Englert} 
\affiliation{Institute for Particle Physics Phenomenology, Department
  of Physics,\\Durham University, DH1 3LE, United Kingdom}
\author{Michael Spannowsky} 
\affiliation{Institute for Particle Physics Phenomenology, Department
  of Physics,\\Durham University, DH1 3LE, United Kingdom}

\preprint{IPPP/12/43}
\preprint{DCPT/12/86}

\maketitle

\section{Introduction}
\label{sec:intro}
The Standard Model (SM) Higgs~\cite{orig} has recently been excluded
at the 95\% confidence level from 129 (127.5) GeV to 539 (600) GeV by
measurements performed by ATLAS (CMS)~\cite{moriond}. In addition, the
Higgs mass bound from LEP2 was raised from 114.4~GeV~\cite{LEP2} to
117.5~GeV by ATLAS. Both ATLAS and CMS have also observed tantalizing
hints for a SM-like Higgs at a mass $m_h\simeq 125~\gev$ with local
significances of 2.5$\sigma$ and 2.8$\sigma$, respectively. In the
same mass region, the D$\slashed{0}$ and CDF collaborations observe an
excess with a local significance of 2.2$\sigma$ for the
combination of their data sets~\cite{tevat}.

Breaking down these results into the individual search channels has
triggered some effort to pin down the properties of the observed
excess in the SM and beyond~\cite{bsm1}. These analyses are the first
steps of a spectroscopy program which targets the properties of a
newly discovered particle if the excess at 125~GeV becomes
statistically significant. Strategies to determine spin- and
${\cal{CP}}$ quantum numbers, and the couplings to fermions and gauge
bosons of a 125~GeV resonance with SM-like cross sections have been
discussed in the literature~\cite{Plehn:2001nj,Lafaye:2009vr}. A
determination of the Higgs self-interaction, however, which is crucial
for a measurement of the symmetry breaking sector remains challenging
in the context of the SM (this can change in BSM
scenarios~\cite{Contino:2012xk}). Even for scenarios where the Higgs
is close to the $h\to W^+W^-$ threshold, statistics at the LHC in
$pp\to hh+X$ is extremely
limited~\cite{Baur:2003gp,Baur:2002qd,Baur:2002rb}, so that
formulating constraints on the Higgs self-coupling requires
end-of-LHC-lifetime statistics if possible at all\footnote{A related
  analysis of Higgs physics at a future linear collider can be found
  in Ref.~\cite{Barger:2003rs}.}.

The Higgs self-coupling in the SM follows from equating out the Higgs
potential after the Higgs doublet is expanded around the electroweak
symmetry breaking vacuum expectation value, $H=(0,v+h)^T/\sqrt{2}$ in
unitary gauge:
\begin{equation}
  \label{eq:higgspot}
  \begin{split}
    V(H^\dagger H) &= 
    \mu^2 H^\dagger H + \eta (H^\dagger H)^2 \\
    &\supset {1\over 2} m_h^2h^2 + \sqrt{ {\eta\over 2}}m_h h^3
    + {\eta\over 4}h^4\,,
  \end{split}
\end{equation}
where $m_h^2=\eta v^2/2$, and $v^2=-\mu^2/\eta$. Since symmetry
breaking in the SM relies on $\mu^2<0$ and $\eta > 0$, the partial
experimental reconstruction of the Higgs potential via the measurement
of the trilinear Higgs vertex and its comparison to SM quantities
({\emph{e.g.}} $2\eta=g^2m_h^2/m_W^2$) is necessary to verify symmetry
breaking due to a SM-like Higgs sector. Strictly speaking, a similar
program needs to be carried out for the quartic Higgs vertex to fully
reconstruct the Higgs potential by measurements. This task is,
however, even more challenging due to an even smaller cross section of
triple Higgs production~\cite{Plehn:2005nk,kauer}.

\begin{figure*}[tbp!]
  \includegraphics[angle=-90,width=0.67\textwidth]{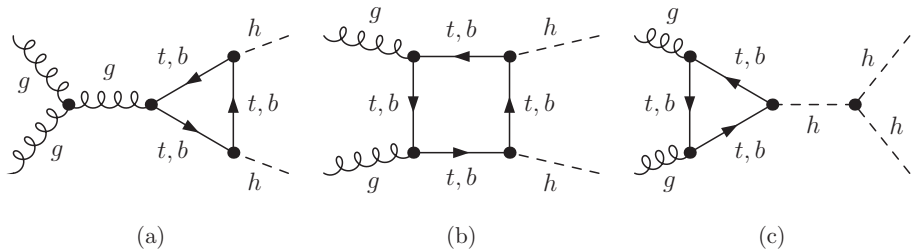}
  \caption{\label{fig:hhgraphs} Sample Feynman graphs contributing to
    $pp\to hh+X$. Graphs of type (a) yield vanishing contributions due
    to color conservation.}
\end{figure*}

A process at hadron colliders which is sensitive to the trilinear
Higgs coupling is the previously mentioned Higgs pair production
$pp\to hh+X$ via gluon fusion. Consequently, this process has already
been studied in the literature in
detail~\cite{Glover:1987nx,Plehn:1996wb,Dawson:1998py,kauer}. From the
known results it is clear that the LHC's potential towards measuring
the trilinear coupling only in a single channel is
insufficient. Combining different Higgs decay final states improves
the situation, but exhausting the entire LHC search potential requires
also discussing the machine's capability to constrain the trilinear
Higgs coupling in different kinematical regimes.

Different properties of signal and background processes in {\it{e.g.}}
boosted final states as compared to unboosted kinematics allows us to
access Higgs decay channels which are impossible to isolate from the
background in a more inclusive search. This has impressively been
demonstrated in the context of subjet-based analysis
techniques~\cite{Butterworth:2008iy} which have proven successful
during the 7~TeV LHC run~\cite{atlasboostedw}. In addition to that,
initial state radiation can be an important effect when considering
energetic final states. A dihiggs system recoiling against a hard
hadronic jet accesses an entirely new kinematical
configuration\footnote{The phenomenology of such configurations can
  also be treated separately from radiative correction contributions
  to $pp\to hh+X$.}, which is characterized by a large dihiggs
invariant mass, but with a potentially smaller Higgs $s$-channel
suppression than encountered in the back-to-back configuration of
$gg\to hh$.

The goal of this paper is to provide a comparative study of the
prospects of the measurement of the trilinear Higgs coupling applying
contemporary simulation and analysis techniques. In the light of
recent LHC measurements, we focus our eventual analyses on
$m_h=125~\gev$. However, we also put this particular mass into the
context of a complete discussion of the sensitivity towards the
trilinear Higgs coupling over the entire Higgs mass range $m_h\lesssim
1~\tev$. As we will see, $m_h\simeq 125~\gev$ is a rather special
case. Since Higgs self-coupling measurements involve end-of-lifetime
luminosities we base our analyses on a center-of-mass energy of
14~TeV.

We begin with a discussion of some general aspects of double Higgs
production, before we review inclusive searches for $m_h=125~\gev$ in
the $pp\to hh+X$ channel in Sec.~\ref{sec:hhi}. We discuss boosted
Higgs final states in $pp\to hh+X$ in Sec.~\ref{sec:hhb} before we
discuss $pp\to hh+j+X$ with the Higgses recoiling against a hard jet
in Sec.~\ref{sec:hhj}. Doing so we investigate the potential
sensitivity at the parton- and signal-level to define an analysis
strategy before we apply it to the fully showered and hadronized final
state. We give our conclusions in Sec.~\ref{sec:summary}.

\section{Higgs pair production at the LHC}
\label{sec:hh}
\subsection{General Remarks}
Inclusive Higgs pair production has already been studied in
Refs.~\cite{Plehn:1996wb,Dawson:1998py,Glover:1987nx,kauer} so we
limit ourselves to the details that are relevant for our analysis.

Higgs pairs are produced at hadron colliders such as the LHC via a
range of partonic subprocesses, the most dominant of which are
depicted in Fig.~\ref{fig:hhgraphs}. An approximation which is often
employed in phenomenological studies is the heavy top quark limit,
which gives rise to effective $ggh$ and $gghh$
interactions~\cite{Shifman:1979eb}
\begin{equation} 
  {\cal{L}_{\mathrm{eff}}} = {1 \over 4}{\alpha_s \over
    3\pi} G^a_{\mu\nu} G^{a\, \mu\nu} \log(1+ h/v)\, ,
\end{equation}
which upon expansion leads to
\begin{equation}
  \label{eq:effth}
  {\cal{L}}\supset +
  {1 \over 4} {\alpha_s \over 3\pi v} G^a_{\mu\nu} G^{a\, \mu\nu} h
  -{1\over 4}  {\alpha_s \over 6 \pi v^2}  G^a_{\mu\nu} G^{a\, \mu\nu} h^2 \,.
\end{equation}
Studying these operators in the $hh+X$ final state should in principle
allow the Higgs self-coupling to be constrained via the relative
contribution of trilinear and quartic interactions to the integrated
cross section. Note that the operators in Eq.~\gl{eq:effth} have
different signs which indicates important interference between the
(nested) three- and four point contributions to $pp\to hh+X$ already
at the effective theory level.

On the other hand, it is known that the effective theory of
Eq.~\gl{eq:effth} insufficiently reproduces all kinematical properties
of the full theory if the interactions are probed at momentum
transfers $Q^2\gtrsim m_t^2$ \cite{Baur:2002rb} and the massive quark
loops are resolved. Since our analysis partly relies on boosted final
states, we need to take into account the full one-loop contribution to
dihiggs production to realistically model the phenomenology.

\subsection{Parton-level considerations}
\label{sec:hhpart}
In order to properly take into account the full dynamics of Higgs pair
production in the SM we have implemented the matrix element that
follows from Fig.~\ref{fig:hhgraphs} in the {\sc{Vbfnlo}}
framework~\cite{vbfnlo} with the help of the
{\sc{FeynArts}}/{\sc{FormCalc}}/{\sc{LoopTools}} packages~\cite{hahn},
with modifications such to include a non-SM trilinear Higgs
coupling\footnote{The signal Monte Carlo code underlying this study is planned to become
part of the next update of {\sc{Vbfnlo}} and is available upon request
until then.
}. Our setup allows us to obtain event files according to the
Les Houches standard~\cite{lhe}, which can be straightforwardly
interfaced to parton showers. Decay correlations are trivially
incorporated due to the spin-0 nature of the SM Higgs boson.

\begin{figure*}[htbp!]
  \includegraphics[height=0.3\textwidth]{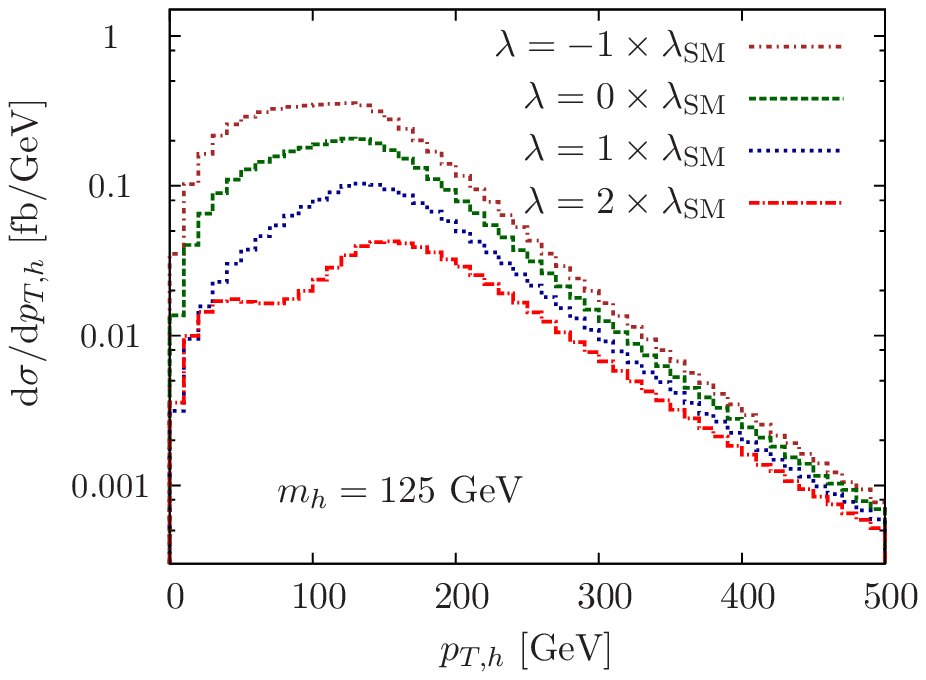}\hspace{1cm}
  \includegraphics[height=0.3\textwidth]{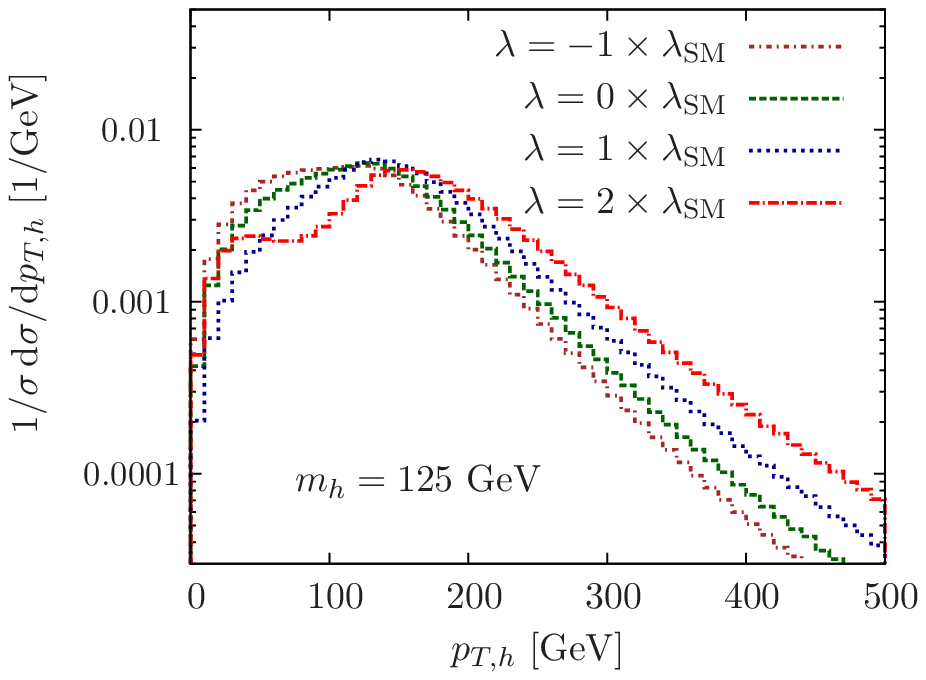}
  \caption{\label{fig:hhprodpt} Comparison of the (normalized)
    $p_{T,h}$ distributions in $pp\to hh+X$ at LO for different
    multiples of the trilinear Higgs coupling~$\lambda$
    ($m_t=172.5~\gev$ and $m_b=4.5~\gev$ using CTEQ6l1 parton
    densities).}
\end{figure*}

\begin{figure}[tbp!]
  \includegraphics[height=0.3\textwidth]{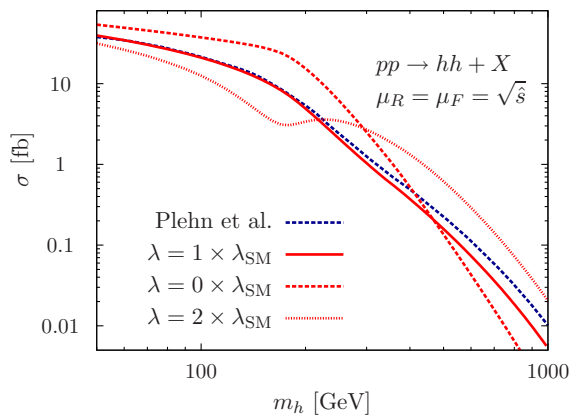}
  \caption{\label{fig:hhprod} Comparison of $pp\to hh+X$ at LO. We
    choose $m_t=175~\gev$ as in Ref.~\cite{Plehn:1996wb}, from which
    we also obtain the dashed blue reference line, and $m_b=4.5~\gev$
    and we use the CTEQ6l1 parton distributions.}
\end{figure}

The resulting inclusive hadronic cross sections are plotted in
Fig.~\ref{fig:hhprod}, where we also show results for non-SM
trilinear couplings, varied around the SM value (see
Eq.~\gl{eq:higgspot})
\begin{equation}
  \label{eq:deflambda}
  \lambda_{{\rm{SM}}}=\sqrt{ {\eta\over 2}}m_h\,.
\end{equation}
Note that choosing a value different from $\lambda_{\rm{SM}}$ does
not yield a meaningful potential in terms of Eq.~\gl{eq:higgspot}, but
allows to constrain $\lambda$ in hypothesis tests using, {\emph{e.g.}},
the CL$_{\rm{s}}$ method \cite{cls}.

We also show the result of Ref.~\cite{Plehn:1996wb} for comparison
and find excellent agreement in total, keeping in mind that the
results of Ref.~\cite{Plehn:1996wb} were obtained using the GRV
parametrizations of parton luminosities~\cite{GRV}, which are
different from the CTEQ6l1~\cite{cteq} set that we employ for the
remainder of this paper\footnote{Using the integration-mode of
  {\sc{FormCalc}}/{\sc{LoopTools}} with the CTEQ6l1 set we obtain
  perfect agreement.}. 

Interference between the different non-zero contributions depicted in
Fig.~\ref{fig:hhprod} becomes obvious for the differently chosen Higgs
self-couplings. We also learn from Fig.~\ref{fig:hhprod} that the
dihiggs cross section has a fairly large dependence on the particular
value of the trilinear coupling for a $m_h=125~\gev$ Higgs boson. The
qualitative Higgs mass dependence for different values of the
trilinear self-coupling in Fig.~\ref{fig:hhprod} is easy to
understand: The Higgs propagator in Fig.~\ref{fig:hhgraphs}~(c) is
always probed off-shell at fairly large invariant masses; this renders
the triangle contributions subdominant compared to the box
contributions of Fig.~\ref{fig:hhgraphs} (b). For Higgs masses close
to the mass of the loop-dominating top quark, we have $s\simeq
4m_t^2$, which results in resonant contributions of the three-point
functions of Fig.~\ref{fig:hhgraphs} (c), well-known from one-loop
$gg\to h$ production~\cite{gfusion}. This ameliorates the $s$-channel
suppression of the trilinear coupling-sensitive triangle graphs and
causes the dependence of the cross section on the trilinear coupling
to become large at around $m_h\lesssim m_t$.

To gain sensitivity beyond total event counts, it is important to
isolate the region of phase space which is most sensitive to
modifications of the trilinear coupling in order to set up an analysis
strategy which targets the trilinear self-coupling most
effectively. At the parton level, there is only a single
phenomenologically relevant observable to $hh$ production, which can
be chosen as the Higgs transverse momentum $p_{T,h}$. In
Fig.~\ref{fig:hhprodpt} we show the differential $p_{T,h}$
distribution for different values of $\lambda$ and $m_h=125~\gev$. The
dip structure for $\lambda>\lambda_{\rm{SM}}$ results again from phase
space regions characterized by $s\sim 4m_t^2$, which are available if
$m_h<m_t$, and the resulting maximally destructive interference with
the box contributions.

\begin{table*}[tbp!]
\begin{tabular}{l|rrr| r | c}
\hline 
& $\xi=0$ & $\xi=1$ & $\xi=2$ & $b\bar b WW$ & ratio to ${\xi=1}$ \\
\hline
cross section before cuts  &  59.48 & 28.34 &  13.36 &  877500 &  $3.2 \cdot 10^{-5}$   \cr
1 isolated lepton  &  7.96 & 3.76 &  1.74  &   254897  & $1.5 \cdot 10^{-5}$  \cr
MET + jet cuts  &  1.54 & 0.85 & 0.44 &  66595.7  &  $1.2 \cdot 10^{-5}$   \cr
hadronic $W$ reconstruction  &  0.59 & 0.33 & 0.17 & 38153.3 &  $0.9 \cdot 10^{-5}$   \cr
kinematic Higgs reconstruction  &  0.028 &   0.017 & 0.007  & 205.1 &  $8.3 \cdot 10^{-5}$   \cr
\hline
\end{tabular}
\caption{\label{tab:bbww}
  Signal and background cross sections in fb for $hh\to b\bar b
  W^+W^-$. The Higgs self-coupling is scaled in multiples of the 
  Standard Model value $\lambda=\xi \times \lambda_\mathrm{SM}$, 
  Eq.~\gl{eq:deflambda}. The background is $b\bar bW^+W^-$ production
  discussed at NLO in Ref.~\cite{Denner:2010jp} ($K\simeq 1.5$).}
\label{tab:cxn2d}
\end{table*}

The above points suffice to give a qualitative assessment of the
prospects of measurements of $\lambda$ in the $pp\to hh+X$ channel:
\begin{itemize}
\item the Higgs bosons from inclusive dihiggs productions are
  naturally boosted $p_{T,h}\gtrsim 100~\gev$,
\item interference leads to an \textit{a priori} $\lambda$-sensitive
  phenomenology for $m_h\simeq 125~\gev$,
\item identical interference effects also cause the bulk of the
  sensitivity to $\lambda$ to follow from configurations with
  $p_{T,h}\sim 100~\gev$, while the $p_{T,h}$ shape at large values
  becomes similar for different values of $\lambda$ due to decoupling
  the triangle contributions at large partonic $\sqrt{\hat s}$,
\item the cross section shows a dependence on the trilinear coupling
  of $\Delta \sigma/\sigma_{\rm{SM}}\simeq 50\%$ when varying
  $\lambda \in [0,2 \lambda_{\rm{SM}}]$.
\end{itemize}

We conclude our parton-level discussion of dihiggs production by
noting that the higher-order corrections~\cite{Dawson:1998py}, which
result in a total $K$ factor of
$\sigma^{\rm{NLO}}/\sigma^{\rm{LO}}\gtrsim 1.85$, result from a large
contribution from real parton emission. This is a characteristic trait
of processes involving color-singlet final states at leading order,
for which plenty of phase space for extra parton emission in addition
to new initial-state parton combinatorics becomes available at
next-to-leading order (NLO). Similar observations have been made for
$pp\to VV+X$, where $V=W^\pm,Z,\gamma$~\cite{dibos}. The discussed
characteristics of $pp\to hh+X$ are therefore not distorted when
including NLO precision, and the parton shower will capture the
characteristic features of the cross section upon normalizing to the
NLO inclusive rate.

\subsection{Inclusive Higgs Pair Searches}
\label{sec:hhi}

To measure $\lambda$ for a 125~GeV Higgs we need to isolate
modifications of $pp\to hh+X$ cross sections around
$\sigma^{\rm{NLO}}(hh+X)=28.4~\fb$ from the Higgses' exclusive decay
channels. Given the small total inclusive cross section\footnote{The
  total inclusive single Higgs production cross section is
  $16.5~\pb$~\cite{Dittmaier:2011ti} for comparison.}, it is clear
that even for $\sqrt{s}=14~\tev$ and a target luminosity of
${\cal{O}}(1000{\rm{fb}}^{-1})$ we need to focus on the Higgs decay
channels with the largest branching ratios to phenomenologically
visible final states to observe $pp\to hh+X$. These are~\cite{hdecay}
$h\to b\bar b$ (59.48\%), $h\to W^+W^+$ (20.78\%), and $h\to \tau\tau$
(6.12\%). The decay $h\to ZZ$ (2.55\%) is limited by the decays of the
$Z$ bosons to the clean leptonic final states $Z\to e^+e^-,\mu^+\mu^-$
(6.67\%) (yielding ${\rm{BR}}(h\to
e^+e^-+\mu^+\mu^-)=0.013\%$). Hadronic $Z$ decay modes can only be
accessed in the boosted regime, which is not feasible for
$m_h=125~\gev$~\cite{boostzz}.

We do not consider the final state $hh\to b\bar b \gamma \gamma$. A
feasibility study for this particular channel was already presented in
Ref.~\cite{Baur:2003gp}. A realistic assessment of the sensitivity in
$b\bar b \gamma \gamma$ depends on a realistic simulation of the
diphoton fake rate due to multijet production, which is the dominant
background to such an analysis, similar to Higgs searches in $h\to
\gamma \gamma$. Details of the photon identification rely on the
detector properties and the event selection approach, and we cannot
address these issues in a realistic fashion. We focus in on $h\to
b\bar b, W^+W^-,\tau^+\tau^-$ in the following.

We generate all (fully showered and hadronized) background samples
with {\sc{Sherpa}}~\cite{Gleisberg:2008ta} or
{\sc{MadEvent}}~\cite{Alwall:2011uj}. The signal events are interfaced
to {\sc{Herwig++}}~\cite{Bahr:2008pv} for showering and hadronization.

\begin{table*}[tbp!]
\begin{tabular}{l|rrr|rrr|c}
\hline 
& $\xi=0$ & $\xi=1$ & $\xi=2$ & $b \bar b b
\bar b$ [QCD] & $b \bar b b \bar b$ [ELW] & $b \bar b b \bar b$ [QCD/ELW] &
ratio to $\xi=1$\\
\hline
cross section before cuts  & 59.48 & 28.42 & 13.36 & 21165  &  16.5 & 160.35 &  $1.3 \cdot 10^{-3}$   \cr
trigger+no leptons  & 17.93 & 10.21 &  5.31 & 5581.2  &  8.0  & 38.85  & $1.8 \cdot 10^{-3}$   \cr
fatjet cuts & 13.73 & 8.23 & 4.50 & 4761.0 & 7.50 & 31.65 & $1.7 \cdot 10^{-3}$   \cr
first Higgs rec + $2b$  & 1.55 & 1.02 & 0.60 & 235.22 & 0.75 & 1.32 & $4.2 \cdot 10^{-3}$   \cr
second Higgs rec + $2b$  & 0.137 & 0.094 & 0.059 & 9.72  &  0.011 & 0.050 & $9.6 \cdot 10^{-3}$   \cr
\hline
\end{tabular}
\caption{\label{tab:bbbb_boosted} 
  Signal and background cross sections in fb for $hh\to b\bar b
  b\bar b$ for boosted kinematics. The Higgs self-coupling is 
  scaled in multiples of the Standard Model value $\lambda=\xi 
  \times \lambda_\mathrm{SM}$, Eq.~\gl{eq:deflambda}. Signal and background
  are normalized to the respective NLO cross sections. NLO $b \bar b b
  \bar b$ production has been provided in Ref.~\cite{Greiner:2011mp}
  (inclusive $K\simeq 1.5$). The mixed QCD+electroweak production and
  the purely electroweak production is currently not know at NLO QCD
  precision, and we therefore use an identical correction factor as for
  the QCD-induced production.
}
\end{table*}

\subsubsection{$hh\to b\bar b b \bar b$}
The exclusive decay to two $b\bar b$ pairs is the most obvious channel
to check for sensitivity due to its large branching ratio for
$m_h=125~\gev$. Using $b$-tagging, it also possible to access the
intermediate invariant Higgs masses, for which the modifications due
to $\lambda\neq \lambda_{\rm{SM}}$ are well-pronounced.

Passing the trigger-level cuts is not a problem for the signal events:
the Higgses are naturally boosted and the $p_T$-ordered $b$ jets
typically pass the staggered cuts on the transverse momentum
$p_{T,j_1}>100~\gev$, $p_{T,j_2}>80~\gev$, $p_{T,j_3}>50~\gev$,
$p_{T,j_4}>40~\gev$. However, as already mentioned, there is only one
relevant scale to dihiggs production, and therefore our options to
compete with the gigantic QCD $pp\to b\bar b b \bar b+X$ background
are highly limited. Note that both Higgs bosons need to be
reconstructed in order to be sensitive to the modifications of the
trilinear coupling.

In total, inclusive dihiggs production with decay to four $b$ quarks
has a signal-over-background ratio $S/B$ which is too bad to be a
suitable search channel, already when we focus only on the QCD-induced
four $b$ background. Hence, this channel is not promising at the
inclusive level and we revise it for boosted kinematics in
Sec.~\ref{sec:hhb}.

\subsubsection{$hh\to b\bar b W^+W^-$}
To gain multiple phenomenological handles to deal with the
contributing backgrounds while preserving a signal rate as large as
possible we focus on $h\to \bar b b$ and $h \to (W \to jj) (W \to \nu
l)$. The contributing background processes are $t \bar t$ and $b\bar b
W^+ W^-$ production and we can employ cuts on missing transverse
energy (MET), lepton identification and $p_T$, and the reconstructed
$W$ resonance to reduce them.

We require exactly one isolated lepton with $p_{T,l} > 10$~GeV in the
central part of the detector $|y|< 2.5$, where isolation means an
hadronic energy deposit of $E_{T,{\rm{had}}} < 0.1\, E_{T,l}$ within a
cone of $R=0.3$ around the identified light-flavor lepton.
In the next step we reconstruct the missing transverse energy (MET)
${\slashed{E}}_T$ from all visible final state objects within the
rapidity coverage $|y|<4.5$ and require ${\slashed{E}}_T>20~\gev$.
Then, we use the anti-kT~\cite{Cacciari:2008gp} algorithm as
implemented in {\sc{FastJet}} \cite{Cacciari:2011ma} (which we use
throughout this paper) to reconstruct jets with $R=0.6$ and
$p_T>40$~GeV, and require at least four jets in $|y|<4.5$. Afterwards
we reconstruct the $W$ boson by looping over all jet pairs. The jet
pair that reconstructs the $W$ mass best within $60~\gev \leq m_{jj}
\leq 100~\gev$ is identified as the $W$ boson, and we subsequently
remove these jets from the event. Analogous to the $W$ reconstruction
we reconstruct the Higgs within $110~\gev \leq m_{jj} \leq
140~\gev$. To reduce the backgrounds and identify the signal
contributions we use a double $b$-tag for the two jets which
reconstruct the Higgs best. We use an efficiency of 60\% with a fake
rate of 2\% in $|y|<2.5$ \cite{atltag}. Thus, if one of the Higgs jets
is outside $|y|\leq 2.5$ our reconstruction fails.

The results of this analysis flow can be found in
Tab.~\ref{tab:bbww}. While the cuts bring down the background by a
factor of $4\times 10^3$, they also reduce the signal by nearly the
same amount ($1.5\times 10^3$). The requirement of two reconstructed
Higgses has a strong effect on the background, however the initial
cross-section (inclusively generated) is simply too large for these
cuts to bring down the $S/B$ for this channel to a level for it to be
useful in constraining the Higgs trilinear coupling.

\begin{table*}[tbp!]
\begin{tabular}{l|rrr|rrr|c}
\hline 
& $\xi=0$ & $\xi=1$ & $\xi=2$ & $b\bar b \tau \tau$  & $b\bar b \tau
\tau$ [ELW] & $b\bar bW^+W^-$ &  ratio to ${\xi=1}$\\
\hline
cross section before cuts  &  59.48 & 28.34 &  13.36 &    67.48    &    $ 8.73 $ &  873000  &  $3.2 \cdot 10^{-5}$   \cr
reconstructed Higgs from $\tau$s  &  4.05 & 1.94 &   0.91 &    2.51  &    $ 1.10 $ &  1507.99 &  $1.9 \cdot 10^{-3}$ \cr
fatjet cuts  &  2.27 & 1.09 &  0.65 &    1.29 &  $0.84$ & 223.21 &  $4.8 \cdot 10^{-3}$ \cr
kinematic Higgs reconstruction $(m_{b\bar b})$  
&  0.41 & 0.26 & 0.15 &    0.104 & $0.047$ & 9.50 &  $2.3 \cdot 10^{-2}$ \cr
Higgs with double $b$-tag  & 0.148  & 0.095 &  0.053 & 0.028 &  $0.020$ & 0.15 & 0.48 \cr
\hline
\end{tabular}
\caption{\label{tab:bbtautau_boosted}
  Signal and background cross sections in fb for $hh\to b\bar b
  \tau^+\tau^-$ for boosted kinematics. The Higgs self-coupling is 
  scaled in multiples of the Standard Model value $\lambda=\xi 
  \times \lambda_\mathrm{SM}$, Eq.~\gl{eq:deflambda}. The background
  comprises $t\bar t$ with decays to $t\to b\tau\nu_{\tau}$, and 
  $b\bar b \tau^+\tau^-$ for pure electroweak and mixed
  QCD-electroweak production, normalized to the respective NLO
  rates. The $b\bar b W^+W^-$ NLO cross sections are provided in
  \cite{Denner:2010jp}  ($K\simeq 1.5$), for the mixed and the purely electroweak
  contributions we infer the corrections from $Zb\bar b$ ($K\simeq
  1.4$) and $ZZ$ ($K\simeq 1.6$) production using {\sc{Mcfm}} \cite{zbb,Campbell:2011bn}. 
}
\end{table*}

\subsection{Boosted Higgs Pair Searches}
\label{sec:hhb}
Moving on to the discussion of boosted final states, we can
potentially gain sensitivity in the dominant Higgs decay modes,
{\emph{i.e.}} in the $b\bar b$ channels \cite{Butterworth:2008iy}. The
downside, however, is that we lose sensitivity to modifications of the
trilinear couplings for harder Higgs bosons along the lines of
Sec.~\ref{sec:hhpart}. Nonetheless, a measurement of the magnitude of
the dihiggs cross section is already an important task in itself.

Recently, in the so-called BDRS analysis \cite{Butterworth:2008iy}, it
has been shown that applying jet substructure techniques on fatjets is
a powerful tool to discriminate boosted electroweak-scale resonances
from large QCD backgrounds.  The BDRS approach proposes to recombine
jets using the Cambridge-Aachen (C/A)
algorithm~\cite{Dokshitzer:1997in,Wobisch:1998wt} with a large cone
size to capture all decay products of the boosted resonance. Then one
works backward through the jet clustering and stops when the
clustering meets a so-called "mass-drop" condition: $m_{j_1} < \mu
m_j$ with $\mu=0.66$ and $\min(p^2_{T,j_1},p^2_{T,j_2})/m_j^2 \Delta
R^2_{j_1,j_2} > y_{\mathrm{cut}}$ using $y_{\mathrm{cut}}=0.09$. If
this condition is not met the softer subjet $j_2$ is removed and the
subjets of $j_1$ are tested for a mass drop. As soon as this condition
is met browsing backward through the cluster history the algorithm
stops. In a step called "filtering" the constituents of the two
subjets which meet the mass drop condition are recombined using the
(C/A) algorithm with $R_{\mathrm{filt}}=\min(0.3,R_{b \bar
  b}/2)$. Only the three hardest filtered subjets are kept to
reconstruct the Higgs boson and the two hardest filtered subjets are
$b$-tagged. The filtering step reduces the active area of the jet
tremendously and makes the Higgs-mass reconstruction largely
insensitive to underlying event and pileup.

For the reconstruction of the boosted Higgs bosons in
Sec.~\ref{sec:bbbb} and Sec.~\ref{sec:bbtautau} we adopt this approach
without modifications. It is worth noting that other techniques,
possibly in combination with the BDRS approach, can improve on the
Higgs reconstruction efficiency and can therefore increase the
sensitivity of the following searches~\cite{Soper:2010xk}.

\subsubsection{$hh\to b\bar b b \bar b$}
\label{sec:bbbb}
As already pointed out, the Higgs bosons are naturally boosted, and
requiring two fatjets subject to BDRS tagging
\cite{Butterworth:2008iy} can improve the very bad $S/B$ in the
conventional $pp\to \bar b b \bar b b+X$ search without losing too
much of the dihiggs signal cross section.

In the analysis, we veto events with light leptons $p_{T,l} > 10$~GeV
in $|y|<2.5$ to reduce $t\bar t$, where the leptons are again assumed
isolated if $E_{T,{\rm{had}}} < 0.1 E_{T,l}$ within $R<0.3$. We need to make
sure that the events we want to isolate pass the trigger level. For
this reason, we recombine final state hadrons to jets with $R=0.4$ and
$p_T>40$~GeV and require at least four jets and the following
staggered cuts: $p_{T,j_1}>100$~GeV, $p_{T,j_2}>70$~GeV,
$p_{T,j_3}>50$~GeV. All jets have to be within detector coverage
$|y|<4.5$.

For the events that pass the trigger cuts, we apply a ``fatjet''
analysis, {\emph{i.e.}} require at least two jets with
$p_{T,j}>150$~GeV and $R=1.5$ in the event. We apply the BDRS approach
to both of these fatjets using $\mu_{\rm{cut}} =0.66$ and
$y_{\rm{cut}}=0.09$. The reconstructed Higgs jets need to
reproduce the Higgs mass within a 20~GeV window: $115~\gev \leq m_h
\leq 135~\gev$, and we additionally require that the two hardest
filtered subjets are $b$-tagged.

We generate the backgrounds with exclusive cuts to make our
cut-analysis efficient, yet inclusive enough to avoid a bias. More
precisely we demand two pairs of $b$ quarks to obey $R_{bb}<1.5$,
$p_T(bb)\geq 100~\gev$, $m(bb)\geq 50~\gev$, while $p_{T,b}\geq
20~\gev $, and $|y_b|\leq 2.5$. The \hbox{(anti-)$b$s} are generated
with $R_{bb}\geq 0.2$.

The results are collected in Tab.~\ref{tab:bbbb_boosted}.  Again,
while the cuts allow an improvement in $S/B$ by an nearly an order of
magnitude, we are still left with a small signal rate on top of a very
large background so that this channel is in the end also not
promising.

\begin{figure*}[tbp!]
  \includegraphics[angle=-90,width=0.67\textwidth]{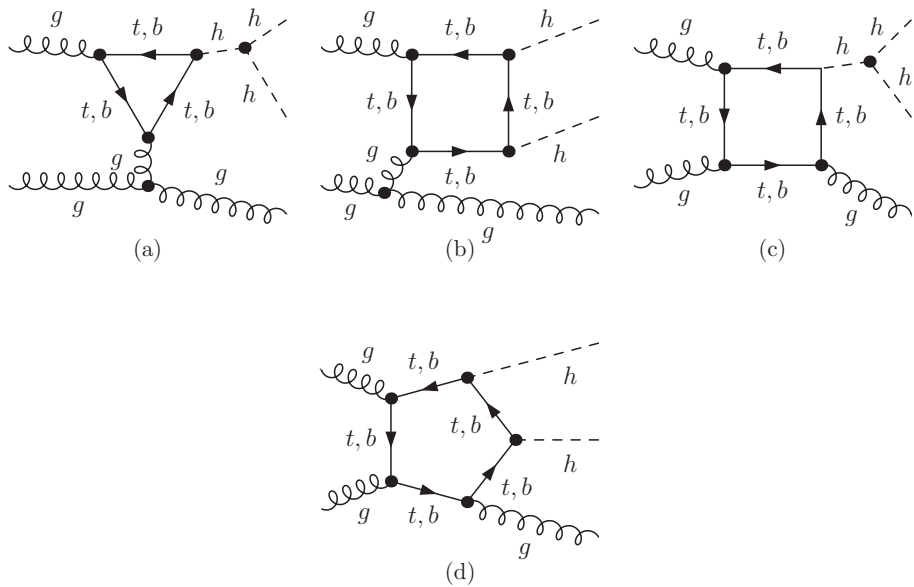}
  \caption{\label{fig:hhjgraphs} Sample Feynman graphs contributing to
    $pp \to hh+j+X$. Not shown are the $qg,\bar q g$ and $q\bar q$
    subprocesses.}
\end{figure*}
\begin{figure}[tbp!]
  \vspace{2cm}
  \includegraphics[width=0.4\textwidth]{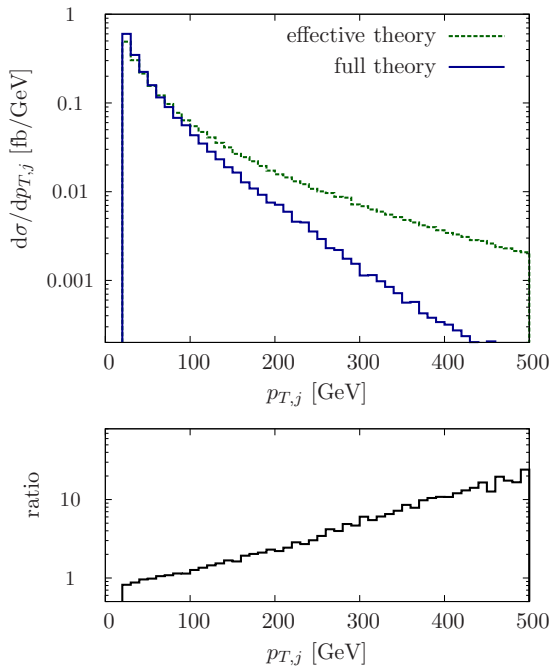}
  \caption{\label{fig:hhjprodpt} Comparison of the leading order
    $p_{T,j}$ spectrum for $pp\to hh+j+X$ production. Shown are
    distributions for the effective interaction (obtained with
    {\sc{MadGraph}} v5~\cite{Alwall:2011uj} via
    {\sc{FeynRules}}~\cite{Christensen:2008py} and
    {\sc{Ufo}}~\cite{Degrande:2011ua}), and the full one-loop matrix
    element calculation. We again choose $m_t=172.5~\gev$ and
    $m_b=4.5~\gev$ using CTEQ6l1 parton densities and
    $\mu_F=\mu_R=p_{T,j}+2m_h$.}
\end{figure}

\begin{figure*}[ph!]
  \includegraphics[height=0.3\textwidth]{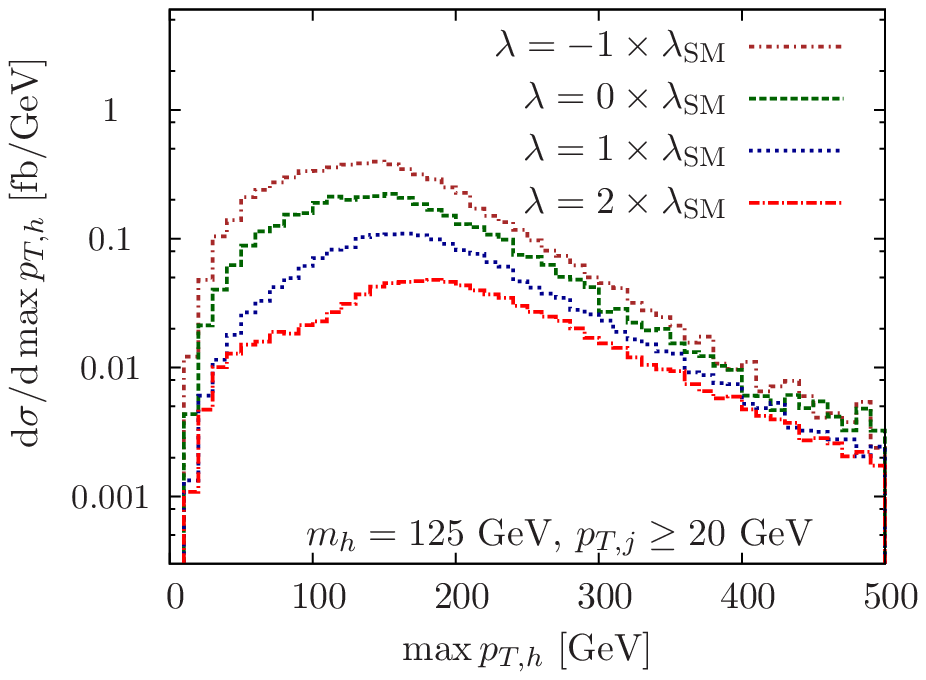}\hspace{1cm}
  \includegraphics[height=0.3\textwidth]{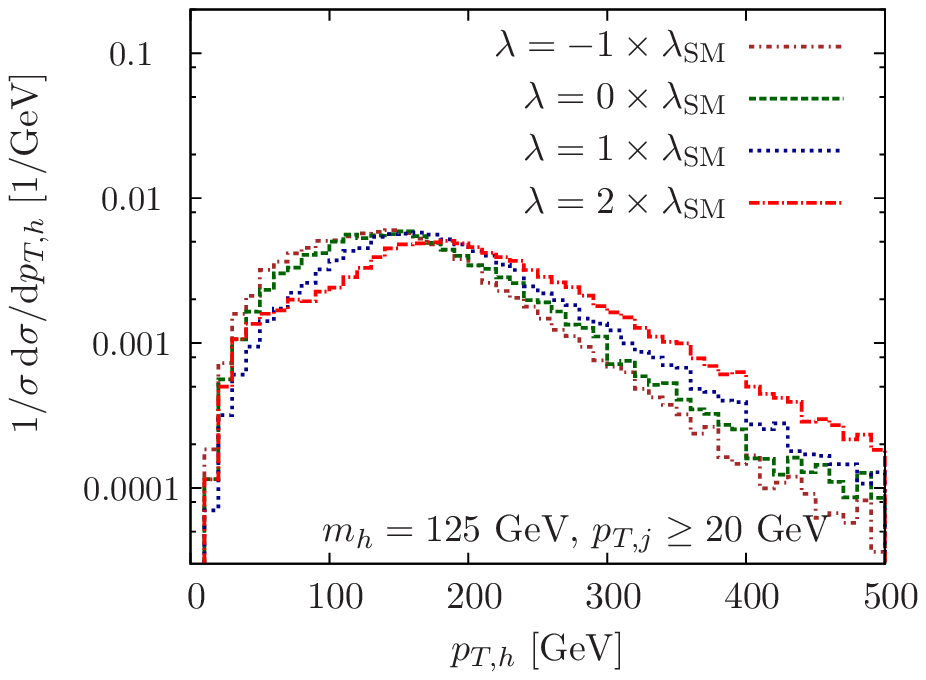}\\[0.4cm]
  \includegraphics[height=0.3\textwidth]{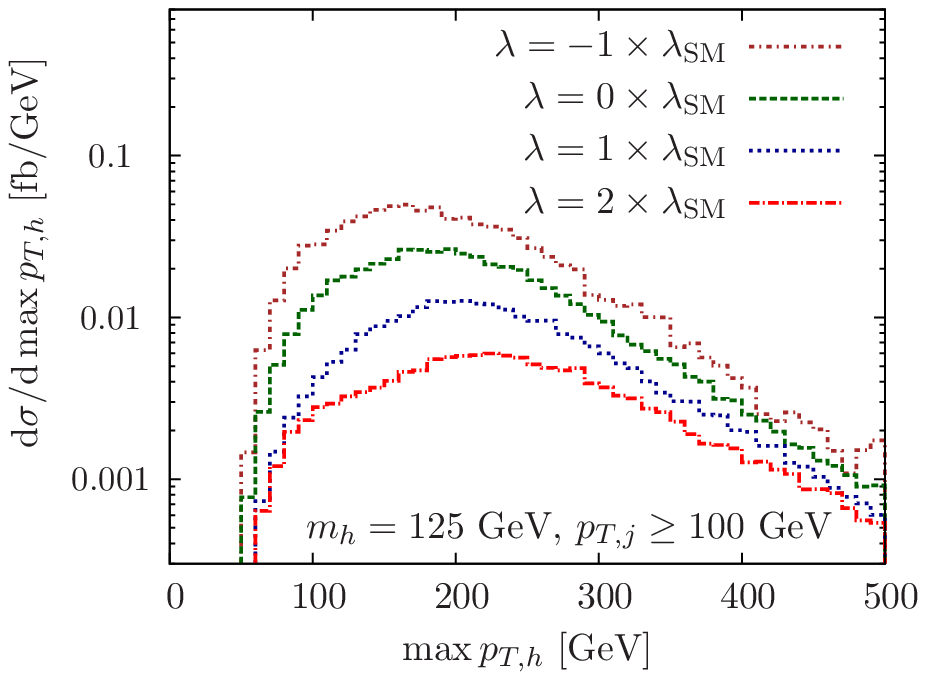}\hspace{1cm}
  \includegraphics[height=0.3\textwidth]{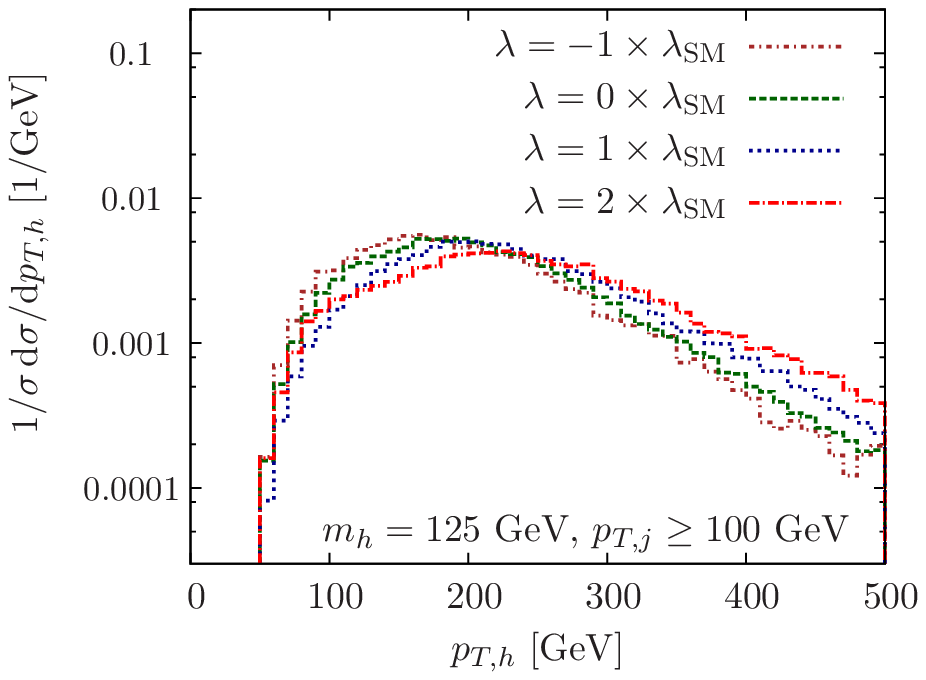} 
  \caption{\label{fig:hhjprodptl} Comparison of the (normalized)
    leading order $\max p_{T,h}$ distributions in $pp\to hh+j+X$ for
    different multiples of the trilinear Higgs coupling $\lambda$
    ($m_t=172.5~\gev$ and $m_b=4.5~\gev$ using CTEQ6l1 parton
    densities), and $p_{T,j}\geq 20~(100)~\gev$ in the upper (lower)
    row, respectively. Factorization and renormalization scales are
    chosen $\mu_F=\mu_R=p_{T,j}+2m_h$.}
\end{figure*}
%
\begin{figure*}[ph!]
  \includegraphics[height=0.3\textwidth]{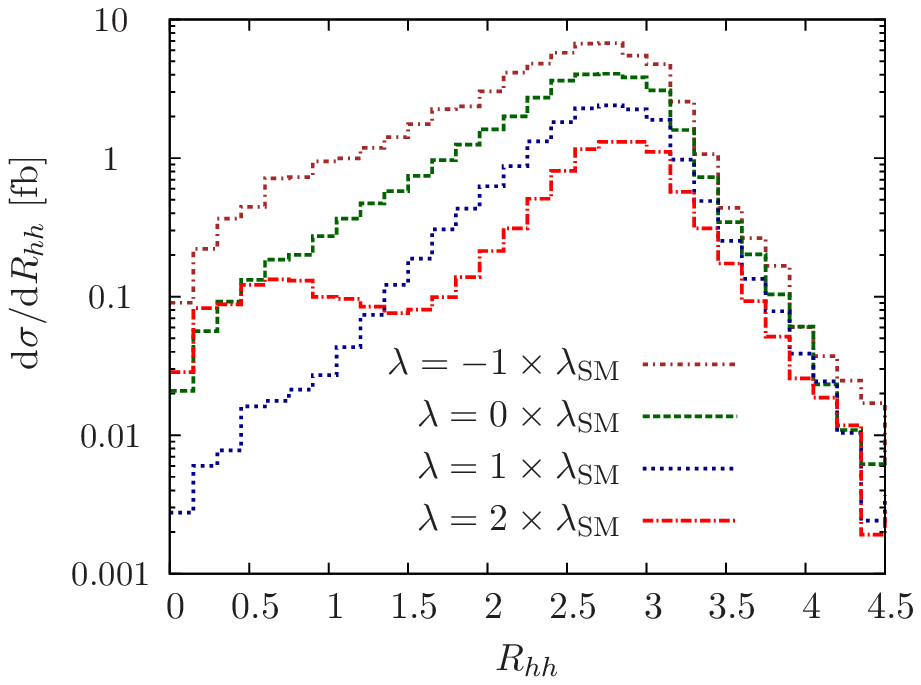}\hspace{1cm}
  \includegraphics[height=0.3\textwidth]{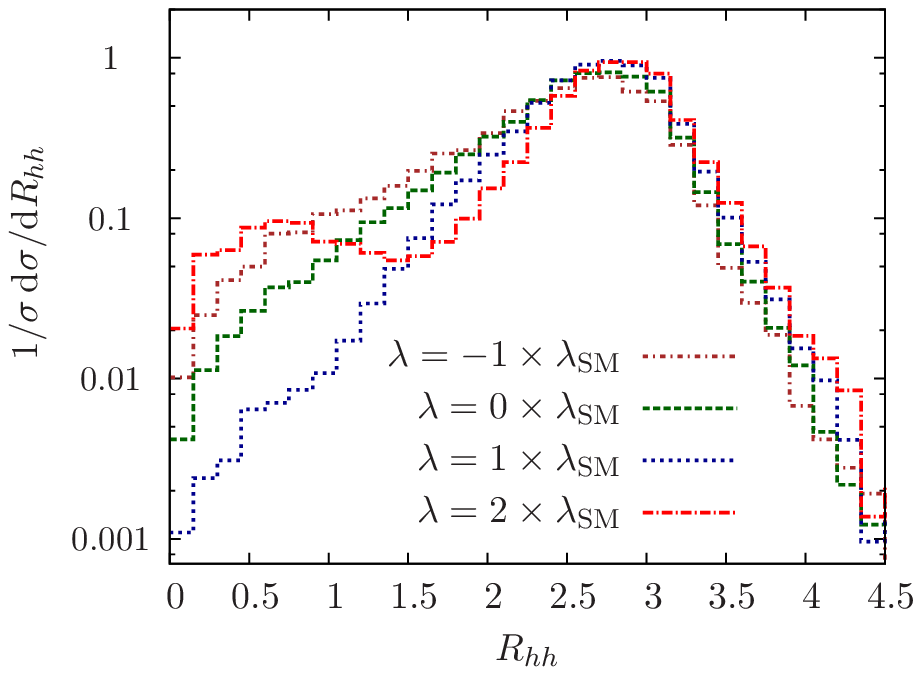} \\[0.4cm]
  \caption{\label{fig:hhjprodRhh} Comparison of the (normalized)
    dihiggs lego plot separation in $pp\to hh+j+X$ at LO for different
    multiples of the trilinear Higgs coupling $\lambda$
    ($m_t=172.5~\gev$ and $m_b=4.5~\gev$ using CTEQ6l1 parton
    densities), and $p_{T,j}\geq 100~\gev$ in the upper (lower) row,
    respectively. Factorization and renormalization scales are chosen
    $\mu_F=\mu_R=p_{T,j}+2m_h$.}
\end{figure*}

\begin{figure*}[tb!]
  \includegraphics[height=0.38\textwidth]{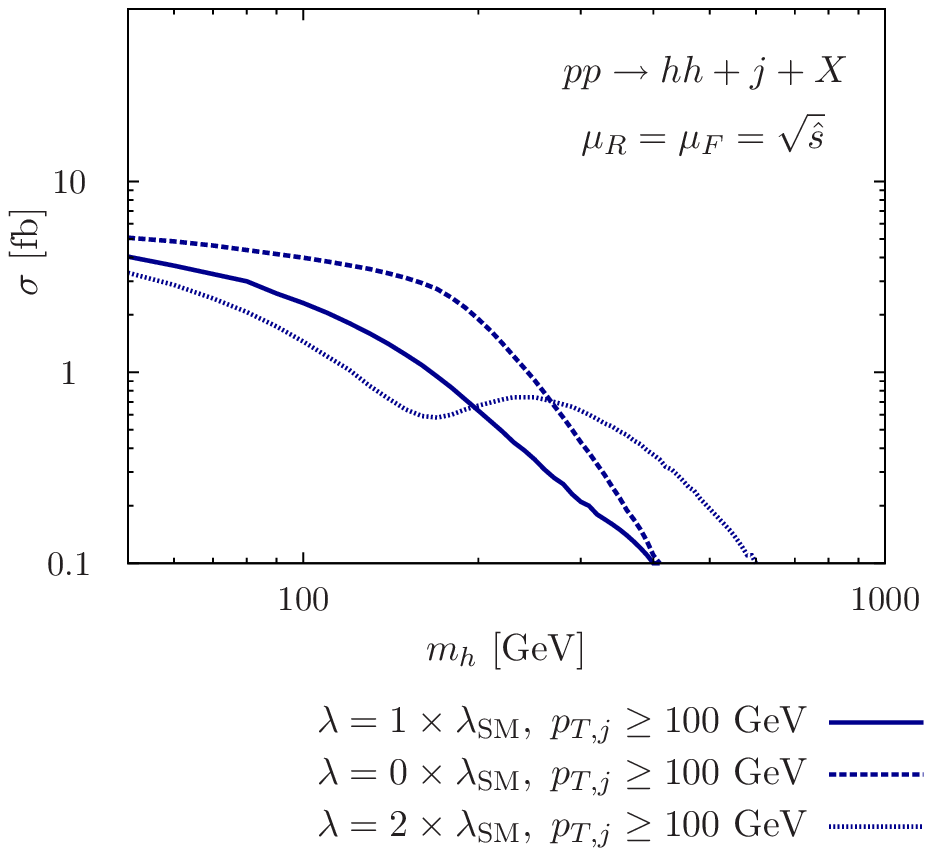}
  \hspace{1cm}
  \includegraphics[height=0.38\textwidth]{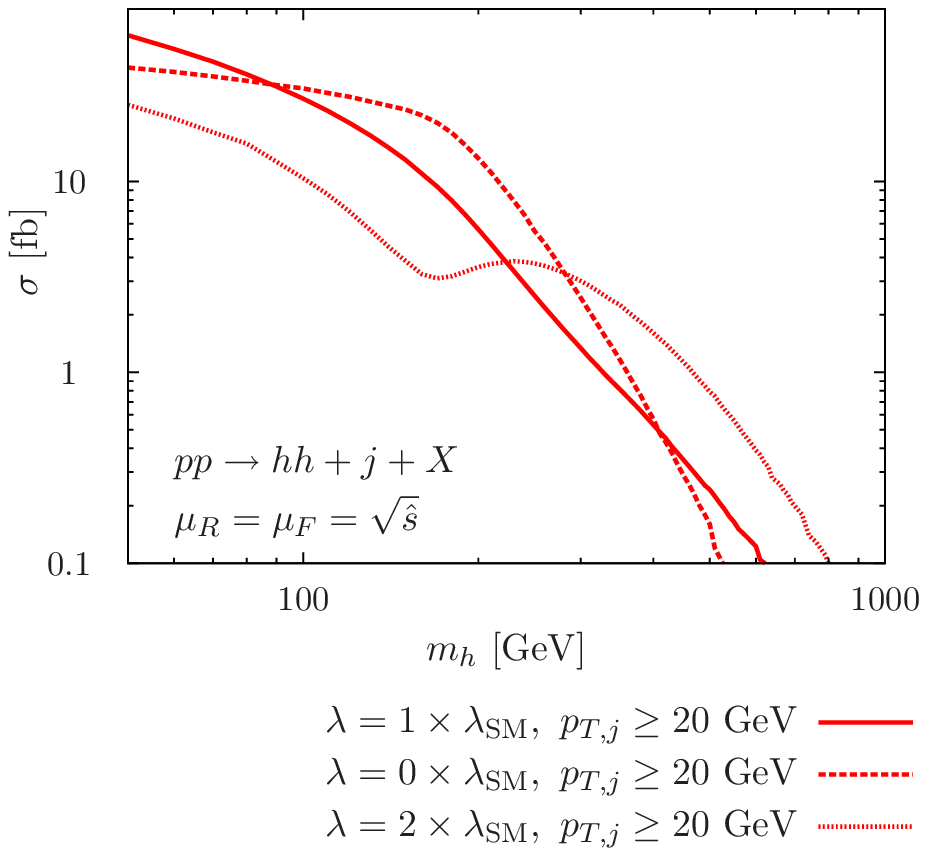}
  \caption{\label{fig:hhjxsexs} Comparison of leading order $pp\to
    hh+j+X$ production cross sections for three different values of
    the Higgs self-coupling and two values of $\min p_T^j$ (we choose
    identical input parameters as for Fig.~\ref{fig:hhjprodpt}).}
\end{figure*}

\subsubsection{$hh\to b\bar b \tau^+ \tau^-$}
\label{sec:bbtautau}
A promising channel is dihiggs production with one Higgs decaying to a
pair of $\tau$ leptons. This decay channel in association with two jets
is one of the main search channels for single light Higgs production
\cite{Rainwater:1998kj,plehn} and has recently been used to put bounds
on Higgs production by {\sc{Cms}} \cite{Chatrchyan:2012vp}. The
reconstruction of $\tau$ leptons is delicate from an experimental point
of view, and current analysis strategies mostly rely on semi-hadronic
$\tau$ pair decays in the context of Higgs searches (see {\emph{e.g.}}
Ref.~\cite{Chatrchyan:2012vp}). The $\tau$ identification is performed
using likelihood methods~\cite{exitau} which do not allow a
straightforward interpretation in terms of rectangular cuts used in
{\emph{e.g.}} Ref.~\cite{plehn}. Consequently, with likelihood $\tau$
taggers unavailable to the public, a reliable and realistic estimate
is hard to obtain. For this reason, we choose a $\tau$ reconstruction
efficiency of 80\% with a negligible fake rate. This is not too
optimistic in the light of the likelihood approaches of
Ref.~\cite{exitau}, bearing in mind that our analyses are based on
end-of-lifetime luminosities, for which we may expect a significant
improved $\tau$ reconstruction when data is better understood.  We choose
a large enough Higgs mass window for the reconstruction, in order to
avoid a too large systematic pollution due to our assumption (in
Ref.~\cite{Chatrchyan:2012vp} CMS quotes a ${\cal{O}}(20\%)$ of the
reconstructed Higgs mass).

In more detail, we require two $\tau$ jets with $p_T\geq 20~\gev$,
reproducing the Higgs mass within 50~GeV, $m_{\tau\tau} = m_h \pm
25~\gev$.  Then we use the C/A algorithm to reconstruct fatjets with
$R=1.5$ and $p_{T,j} > 150$~GeV and require at least one fatjet in the
event. Thereby we demand the fatjets to be sufficiently isolated from
the $\tau$s. We subsequently apply the BDRS approach to the fatjet
with $\mu_{\rm{cut}} = 0.66$ and $y_{\rm{cut}} = 0.09$. The two
hardest filtered subjets need to pass $b$ tags and the reconstructed
Higgs jet has to be in $m_h \pm 10$~GeV. $B$-tagging is performed for
$|y|<2.5$ and we assume an efficiency of 70\% and a fake rate of 1\%
following Ref.~\cite{giac}.

We generate the $b\bar b \tau \tau$ and pure electroweak $b\bar b \tau
\tau$ backgrounds with exclusive cuts to make our cut-analysis
reasonably efficient , yet inclusive enough to avoid a bias. More
precisely we demand the two $b$ quarks to obey $R_{bb}<1.5$,
$p_T(bb,\tau\tau)\geq 100~\gev$, $m(bb,\tau\tau)\geq 50~\gev$, while
$p_{T,b,\tau}\geq 20~\gev $, and $|y_{b,\tau}|\leq 2.5$. The $b$s and
$\tau$s are generated with $R_{bb,\tau\tau}\geq 0.2$. On the other
hand, the $ b \bar b W^- W^+$ sample is generated inclusively, and is
the same sample used in the unboosted $b \bar b W^- W^+$ analysis in
the previous section.

The results are shown in Tab.~\ref{tab:bbtautau_boosted}.  The initial
background cross-section looks very large due to it being inclusively
generated. However, once we take into account the small branching
ratio of $W \to \tau \nu$ this drops dramatically. After requiring two
$b$-tagged jets which reconstruct the Higgs mass we are left with an
$S/B$ of nearly half for the $\xi=1$ case (and nearly one in for
$\xi=0$). The cross-section is also reasonable, corresponding to 95
events for 1000 inverse femtobarns of luminosity. This channel is
hence very promising indeed.

\section{Higgs pair production in association with a hard hadronic
  jet}
\label{sec:hhj}

\subsection{Parton-Level considerations}
\label{sec:hhjpart}

The qualitatively poor agreement of the effective theory of
Eq.~\gl{eq:effth} with the full theory persists if additional jet
radiation is included. Naively we could have expected that accessing
smaller invariant masses in the Higgs system due to significant
initial state radiation might result in a better agreement
with the effective theory of Eq.~\gl{eq:effth}. However, especially for hard
jet emission, which allows the Higgs pairs to access large invariant
masses in a new collinear kinematical configuration compared to $pp\to
hh+X$, the disagreement of full and effective theories is large
(Fig.~\ref{fig:hhjprodpt}).

Given these shortcomings of the effective theory, we implement the
full matrix element in the {\sc{Vbfnlo}} framework using
{\sc{FeynArts}}/{\sc{FormCalc}}/{\sc{LoopTools}}. We have checked our
phase space implementation for the effective theory's matrix element
against {\sc{MadEvent}}. Some of the contributing Feynman graphs to
the dominant $gg$-initiated subprocess are shown in
Fig.~\ref{fig:hhjgraphs}; note that again only a subset of the
contributing diagrams is sensitive to non-standard $hhh$
couplings. Interference between these and the remaining contributions
is again obvious from Fig.~\ref{fig:hhjxsexs} especially at around
$m_h\lesssim m_t$, which can again be explained along the lines of
Sec.~\ref{sec:hhpart}.

In comparison to $pp\to hh+X$, we find sizably larger dependence on
$\lambda$ of the total cross section, Fig~\ref{fig:hhjxsexs}. For
$p_{T,j}\geq 20~\gev$ we have $\Delta\sigma/\sigma_{\rm{SM}}\simeq
100\%$ for a variation $0\leq\lambda \leq 2\lambda_{\rm{SM}}$. This is
due to the larger available phase space for the dihiggs system. The
intermediate $s$ channel Higgs in Fig.~\ref{fig:hhjgraphs} (a),~(c) is
probed at smaller values compared to Fig.~\ref{fig:hhgraphs} (c),
suppression is ameliorated and (destructive) interference becomes more
pronounced.

With a dihiggs system that becomes less back-to-back for increasingly
harder jet emission, the characteristic dip structure encountered in
the $p_{T,h}$ spectrum of $pp\to hh+X$ is washed out
(Fig.~\ref{fig:hhjprodptl}). Characteristic imprints can still be
observed in the dihiggs invariant mass or, equivalently, in the
dihiggs separation in the azimuthal-angle---pseudorapidity plane,
Fig.~\ref{fig:hhjprodRhh}.

Let us summarize a few points relevant to the analysis of
$pp\to b\bar b b\bar b+j+X$ production for $p_{T,j}\gtrsim 100$~GeV:
\begin{itemize}
\item The dihiggs+jet cross section has a comparably large dependence
  on the value of the trilinear couplings as compared to $pp\to hh+X$
  ($\Delta \sigma /\sigma_{\rm{SM}}\simeq 45\%$ when varying $\lambda
  \in [0,2\lambda_{\rm{SM}}]$),
\item the sensitivity to non-standard values of the trilinear coupling
  arises from phase space configurations where the two Higgs bosons
  are close to each other in the central part of the detector,
  {\emph{i.e.}} for rather small values of the invariant masses,
\item as a consequence, the hadronic Higgs decay products are likely
  to overlap, and to fully reconstruct the busy $hh$ decay system we
  need to rely on jet-substructure techniques.
\end{itemize}

\begin{table*}[tb!]
\begin{tabular}{l|rrr|rrr|c}
\hline 
& $\xi=0$ & $\xi=1$ & $\xi=2$ & $b \bar b b \bar b j$ [QCD] & $b \bar b b
\bar b j$ [QCD/ELW] & $b \bar b b \bar b j$ [ELW]  & ratio to ${\xi=1}$\\
\hline
cross section before cuts                             &  6.45 & 3.24 &   1.81 &  29400 &  513.36 & 10.0 & $1.1 \cdot 10^{-4}$   \cr
trigger+fatjet cuts                                                     & 1.82  & 1.08 &   0.69 &     10579.8 &  211.04 & 4.16 & $1.0 \cdot 10^{-4}$   \cr
first kinematic Higgs rec (new tagger) + $2b$                  &  0.30 & 0.20 &   0.13 &    331.84 &       10.82 & 0.54 & $0.5 \cdot 10^{-3}$   \cr
sec kinematic Higgs rec + $2b$                  &  0.09 & 0.059 &0.039 &      54.1 &        2.46 & 0.066 & $1.0 \cdot 10^{-3}$   \cr
invariant mass + $p_{T,j}$ cut                                                &  0.049 & 0.031 &  0.022 &     36.06 &         0.92 & 0.030 & $0.9 \cdot 10^{-3}$   \cr
\hline
\end{tabular}
\caption{ \label{tab:bbbbj_boosted}
  Signal and background cross sections in fb for $hh+j\to b\bar b
  b\bar b+j$ for boosted kinematics. The Higgs self-coupling is 
  scaled in multiples of the Standard Model value $\lambda=\xi 
  \times \lambda_\mathrm{SM}$, Eq.~\gl{eq:deflambda}. None of the
  contributing backgrounds' normalization is known to NLO QCD
  precision. We therefore include a conservative correction factor of $K=2$.}
\end{table*}

Let us again comment on the impact of higher order QCD contributions.
A full NLO QCD computation for $pp\to hh+j+X$ is yet missing, but most
$pp\to VV+j+X$ ($V=W^\pm,Z,\gamma$) production cross sections, which
have similar properties from a QCD point of view, are known to NLO QCD
precision~\cite{vvj}. Also, the NLO QCD cross sections for $pp\to
Vh+j+X$ $(V=W^\pm,Z)$ have been provided in Ref.~\cite{hvj}. Given
that the QCD sector is largely agnostic about the matrix elements'
precise electroweak properties (taken apart the partonic composition
of the initial state), it is not a big surprise that all these
production cross sections exhibit a rather similar phenomenology at
NLO QCD. The total inclusive $K$ factors range around $K\sim 1.3$ and
result from unsuppressed parton emission. It is hence reasonable to
expect the QCD corrections to $pp\to hh+j+X$ to be of similar size,
and parton shower Monte Carlo programs to reasonably reproduce the
dominant kinematical properties. 

For the remainder of this paper we do not include the weak boson
fusion component \cite{figy} to one-jet inclusive production. This the
second largest contribution to inclusive dihiggs production, but it is
still smaller than $hh+j$ production from gluon fusion in the phase
space region we are interested in. For $\max p_{T,j}\geq 80~\gev$ we
have $\sigma_{\rm{WBF}}(hh+2j)\simeq 0.5~\fb$ in the SM, so this
amounts to a ${\cal{O}}(+10\%)$ correction to our inclusive signal
estimate (well inside the perturbative uncertainty of $pp\to hh+j+X$).

\subsection{Boosted Higgs searches in association with a jet}

From Fig.~\ref{fig:hhjprodptl} we see that the Higgs bosons are again
naturally boosted. Events of this signature possess an hadronically
more active final state. The pure BDRS approach works very well if
there is no other hard radiation inside the fatjet except the decay
products of the Higgs boson. Here it is likely that the additional
hard jet ends up in one of the fatjets challenging a good
reconstruction of the Higgs. Therefore, we modify the tagger similar
to the Higgs tagger in \cite{tth}: The last clustering of the jet $j$
is undone, giving two subjets $j_1,j_2$, ordered such that $m_{j_1} >
m_{j_2}$. If $m_{j_1} > 0.8 m_j$ we discard $j_2$ and keep $j_1$,
otherwise both $j_1$ and $j_2$ are kept. For each subjet $j_i$ that is
kept, we either add it to the list of relevant substructures (if $m_i
< 30$~GeV) or further decompose it recursively. After performing this
declustering procedure (we do not stop after observing a
mass drop but continue to decluster the jets until we obtain a set of
hard subjets inside the fat jet) we recombine the constituents of
every two-subjet combination with the C/A algorithm using
$R_\mathrm{filt} = \min(0.3,\Delta R_{j_1,j_2}/2)$. For each
combination we keep the three hardest filtered subjets and call it a
Higgs candidate. The two hardest filtered subjets of the Higgs
candidate with the mass closest to the true Higgs mass of $125$~GeV we
require to be $b$-tagged.  We find that this tagger recovers roughly
$40\%$ more of the signal events while keeping $S/B$ constant.

\begin{table*}[tb!]
\begin{tabular}{l|rrr|rrr|c}
\hline 
& $\xi=0$ & $\xi=1$ & $\xi=2$ & $b \bar b \tau^+ \tau^-j$ & $b \bar b
\tau^+ \tau^-j$ [ELW] & $t \bar t j$  & ratio to ${\xi=1}$ \\
\hline
cross section before cuts  & 6.45  & 3.24  & 1.81 & 66.0 & 1.67 & 106.7 &  $1.9 \cdot 10^{-2}$   \\
2 $\tau$s                                & 0.44 & 0.22  & 0.12 & 37.0 & 0.94 & 7.44 &   $4.8 \cdot 10^{-3}$   \\
Higgs rec. from taus + fatjet cuts                         & 0.29 & 0.16  & 0.10 & 2.00 & 0.150 & 0.947 &   $5.1 \cdot 10^{-2}$   \\
kinematic Higgs rec.         & 0.07 & 0.04  & 0.02 & 0.042 & 0.018 &  0.093 &   $0.26$   \\
$2b$ + $hh$ invariant mass + $p_{T,j}$ cut                          & 0.010 & 0.006  & 0.004 & $<$0.0001 & 0.0022 & 0.0014 &  $1.54$   \\
\hline
\end{tabular}
\caption{\label{tab:bbtautauj_boosted}
  Signal and background cross sections in fb for $hh+j\to b\bar b
  \tau^+\tau^-+j$ for boosted kinematics. The Higgs self-coupling is 
  scaled in multiples of the Standard Model value $\lambda=\xi 
  \times \lambda_\mathrm{SM}$, Eq.~\gl{eq:deflambda}. The QCD
  corrections to $t\bar t+j$ have been discussed in
  Ref.~\cite{Melnikov:2010iu} ($K\simeq 1.1$). For the pure
  electroweak production we take the results of \cite{vvj} as a
  reference value ($K\simeq 1.3$). The corrections to mixed production
  are unknown and we conservatively use a total inclusive QCD
  correction $K=2$.
}
\end{table*}

\subsubsection{$hh+j\to b\bar b b\bar b+j$}
\label{sec:bbbbj}

We proceed in a similar manner to the analysis outlined in
Sec.~\ref{sec:hhb}, but with modifications of the Higgs tagger in
order to preserve a larger signal cross section. 

We generate the backgrounds with the following parton-level cuts, yet
inclusive enough to avoid a bias. We require that two $bb$
combinations obey $p_T(bb)\geq 100~\gev$ and $m(bb)\geq 100~\gev$,
while $|y_b|\leq 2.5$ and $p_{T,b}\geq 20~\gev$. The $b$s are
separated by $R_{bb}\geq 0.2$.  The additional jet is generated with
$p_T\geq 80~\gev$ in $|y_j|\leq 4.5$ and is separated from the $b$s by
$\Delta R \geq 0.7 $. Signal events are generated with $p_{T,j}\geq
80~\gev$.

In the analysis, we again veto events with isolated leptons in
$|y|<2.5$ for $p_{T,l} > 10$~GeV and $E_{T,\rm{{had}}} < 0.1\,
E_{T,l}$ within $R<0.3$. To assure the trigger requirements we ask for
at least five jets and the following staggered cuts on the transverse
momentum: $p_{T,j_1}>120$~GeV, $p_{T,j_2}>100$~GeV,
$p_{T,j_3}>70$~GeV, $p_{T,j_{4}}>40$~GeV. All jets have to fall inside
the detector $|y|<4.5$. The events which pass these trigger criteria
are again analyzed in a subjet approach: We reconstruct fatjets with
$p_{T,j}>150$~GeV and $R=1.5$. At least one fatjet has to be present
which fulfills the mass-drop condition~\cite{Butterworth:2008iy} with
an invariant mass of $m_j> 110$~GeV.  The reconstructed Higgs has to
be double $b$-tagged and have $p_{T,h} > 150$~GeV.

The hardest of these fatjets is declustered with a tagger which is
inspired by the Higgs tagger of Ref.~\cite{tth}. The reconstructed
Higgs has to be double $b$-tagged with $p_{T,h} > 150$~GeV. We
subsequently remove the reconstructed Higgs' constituents from the
event.

The remaining constituents are reclustered using the anti-kT algorithm
with $R=0.4$ and $p_{T,j} > 30$~GeV. The two jets which reconstruct
the Higgs mass best within $m_h=125 \pm 10$~GeV are $b$-tagged with
60\% tagging efficiency and 2\% fake rate; $b$-tagging is again
performed in $|y|<2.5$. Then, the two jets are again removed from the
event. For the two reconstructed Higgs jets we require an invariant
mass $(p_{H_1} + p_{H_2})^2 > 400^2~\gev^2$ and the last remaining jet
needs to be hard $p_T>100$~GeV. In total, this corresponds to a signal
signature discussed in Sec.~\ref{sec:hhjpart}.

The results of this analysis flow can be found in
Tab.~\ref{tab:bbbbj_boosted}. For the numbers quoted there it is clear
that our cut setup serves to highly reduce the contributing
backgrounds, but the QCD-induced cross sections have a too large
initial value. In the QCD background, while the $b\bar b b \bar b$ was
predominantly gluon initiated, the $b\bar b b \bar b + j$ receives
large contributions from $qg$ initial states, leading to final states
with a large invariant mass. This in turn increases the amount of
background in searches for boosted resonances.  The QCD backgrounds
can be reduced by a factor $\mathcal{O}(1000)$ while the signal rate
is decreased by a factor $\sim 100$. In total, this does not leave a
large enough $S/B$ to be relevant from the point of systematics, and
we conclude that $pp\to hh+j\to b\bar b b\bar b+j$ is not a sensitive
channel.

\subsubsection{$hhj\to b\bar b \tau^+ \tau^- j$}

We follow closely the steps described in Sec. \ref{sec:bbtautau} and
Sec. \ref{sec:bbbbj}.  

We generate the backgrounds with the following parton-level cuts to
have a reasonably efficient analysis, yet inclusive enough to avoid a
bias. We require $p_T(b\bar b,\tau\tau)\geq 100~\gev$ and
$m(bb,\tau\tau)\geq 90~\gev$ ($100~\gev$ in case of $t\bar t+j$),
while $|y_{b,\tau}|\leq 2.5$ and $p_{T,b,\tau}\geq 20~\gev$. The $b$s
and $\tau$s are separated by $R_{bb,\tau\tau}\geq 0.2$.  The
additional jet is generated with $p_T\geq 80~\gev$ in $|y_j|\leq 4.5$
and is separated from the $b$s by $\Delta R \geq 0.7 $. Signal events
are generated with $p_{T,j}\geq 80~\gev$.

We require exactly two $\tau$ jets in an event in $|y_{\tau}|<2.5$
with $p_T\geq 20~\gev$ and assume an identification efficiency of $80
\%$ each.  The $\tau$s have to reconstruct to an invariant mass of
$m_h \pm 25$~GeV. Then we use the C/A algorithm to reconstruct fatjets
with $R=1.5$ and $p_{T,j} > 150$~GeV and require at least 1 fatjet in
the event which is sufficiently isolated from the $\tau$s. Then we
apply the Higgs tagger described in Sec.~\ref{sec:hhb} and require the
reconstructed Higgs jet have a mass of $m_h \pm 10$~GeV and $p_{T,H} >
150$~GeV. To suppress the large $t\bar t$ background we reject events
where the invariant mass of the two reconstructed Higgs bosons is
below $400$~GeV. After removing the constituents of the reconstructed
Higgs bosons from the final state we cluster the remaining final state
constituents using the anti-kT algorithm $R=0.4$ and
$p_{T,j}>30$~GeV. Finally, we require at least one jet with
$p_T>100$~GeV.

We find that these cuts can suppress the backgrounds confidently as
long as the $\tau$ fake rate is sufficiently small. Due the large
invariant mass of the final state, several high-$p_T$ jets and
possibly leptons from the $\tau$ decays we expect that these events
can be triggered on easily. The full analysis flow can be found in
Tab.~\ref{tab:bbtautauj_boosted}. The initial background contributions
are significantly lower, as this final state does not have a dominant
purely QCD-induced component. In total we end up with an estimate on
$S/B\simeq 1.5$. This means that with a target luminosity of
$1000~{\rm{fb}}^{-1}$, constraints can be put on $\lambda$ in this
channel.

\section{Summary}
\label{sec:summary}

We have studied the prospects to constrain the trilinear Higgs
coupling by direct measurements at the LHC in several channels,
focussing on $m_h =125$~GeV. This is also the mass region which is
preferred by electroweak precision data, and where we currently
observe excesses in data both at the LHC and the Tevatron. Depending
on the particular decay channel, we find a promising
signal-to-background ratio at the price of a very small event rate.

Higgs self-coupling measurements for a SM Higgs in this particular
mass range are typically afflicted with large backgrounds,
so that achieving maximal sensitivity requires the combination of as
many channels as possible. For dedicated selection cuts we obtain
signal cross sections in Higgs pair production of the order of 0.01 to
0.1~fb and measurements will therefore involve large data sets of the
14 TeV run with a good understanding of the involved experimental
systematics.

Searches for unboosted kinematics of the Higgs bosons do not allow any
constraint on the trilinear coupling or total cross-section to be
made.  However, requiring the two Higgses to be boosted and applying
subjet methods to boosted $pp\to hh+X$ and $pp\to hh+j+X$ production,
we find a sensitive $S/B$ particularly for final states involving
decays into $\tau$s.  A necessary condition for sensitivity in these
channels is a sufficiently good $\tau$ reconstruction, but more
importantly, a small fake rate. Unfortunately, while boosting the
Higgses increases $S/B$, it leads us into a region of phase space which
lacks sensitivity to the trilinear coupling.

In addition to inclusive dihiggs production we find that dihiggs
production in association with a hard jet shows an improved
sensitivity to the trilinear Higgs coupling. However to exploit this
scenario still requires the use of boosted techniques which require
thorough evaluation on data.

Assuming the efficiency for $\tau$-tagging and the hadronic Higgs
reconstruction as outlined in this work are confirmed using data, the
$b\bar b\tau^+\tau^-$ and $b\bar b \tau^+\tau^-+j$ channels can be
used to constrain the Higgs self-coupling in the SM at the LHC with a
data set of several hundred inverse femtobarns. The analysis
strategies developed in this paper will also help to improve bounds on
dihiggs production in scenarios with strong electroweak symmetry
breaking and related models, which also predict enhanced dihiggs
production cross sections \cite{Contino:2012xk,compos}.

\section*{Acknowledgements}
CE acknowledges funding by the Durham International Junior Research
Fellowship scheme and helpful conversations with P.~M.~Zerwas. CE also
thanks the CERN theory group, and MJD the Bonn theory group, for
hospitality during the time when this work was completed. CE thanks
Christophe Grojean, Li Lin Yang and Jos\'e Zurita for comments on the
manuscript and for discussions during the CERN BSM Summer
Institute. We cordially thank the members of the Institute for
Particle Physics Phenomenology for their patience during the time when
we were occupying the entire cluster, and especially Mike Johnson and
Peter Richardson for the cluster-support.

\end{document}